\documentclass[12pt]{article}
\usepackage{graphicx}
\usepackage{amsmath}
\usepackage{bm,latexsym,amsmath,amssymb,amsfonts,mathrsfs}
\def\dd{\mathrm{d}}
\def\mcA{\mathcal{A}}
\def\mcP{\mathcal{P}}
\def\ff{{\rm f}}
\def\em{{\rm em}}
\def\inf{{\rm inf}}
\def\tot{{\rm tot}}
\def\obs{{\rm obs}}
\def\mi{{\rm min}}

\def\Mpl{M_{\rm Pl}}
\def\GeV{{\rm GeV}}

\newcommand\pubnumber{IPMU14-0018}
\newcommand\pubdate{January 30, 2014}

\def\IPMU{ Kavli Institute for the Physics and Mathematics of the Universe (Kavli IPMU),\\ 
TODIAS, the University of Tokyo, 5-1-5 Kashiwanoha, Kashiwa, 277-8583, Japan}
\def\UTokyo{Department of Physics, University of Tokyo, Bunkyo-ku 113-0033, Japan}
\def\support{\footnote{tomohiro.fujita@ipmu.jp}}

\def\Title#1{\begin{center} {\Large #1 } \end{center}}
\def\Author#1{\begin{center}{ \sc #1} \end{center}}
\def\Address#1{\begin{center}{ \it #1} \end{center}}

\newcommand\pubblock{\rightline{\begin{tabular}{l} \pubnumber\\
         \pubdate  \end{tabular}}}
\newenvironment{Abstract}{\begin{quotation}  }{\end{quotation}}
\newenvironment{Presented}{\begin{quotation} \begin{center} 
             PRESENTED AT\end{center}\bigskip 
      \begin{center}\begin{large}}{\end{large}\end{center} \end{quotation}}
\def\Acknowledgements{\bigskip  \bigskip \begin{center} \begin{large}
             \bf ACKNOWLEDGEMENTS \end{large}\end{center}}

\def\beq{\begin{equation}}
\def\eeq#1{\label{#1}\end{equation}}
\def\eeqn{\end{equation}}

\def\beqa{\begin{eqnarray}}
\def\eeqa#1{\label{#1}\end{eqnarray}}
\def\eeqan{\end{eqnarray}}

\let\bar=\overbar

\def\Dslash{\not{\hbox{\kern-4pt $D$}}}
\def\dslash{\not{\hbox{\kern-2pt $\del$}}}

\def\msb{{\bar{\ssstyle M \kern -1pt S}}}

\begin{document}

\begin{titlepage}
\pubblock

\vfill
\Title{Void magnetic field and
its primordial origin in inflation}
\vfill
\Author{ Tomohiro Fujita\support}
\Address{\IPMU. \UTokyo.}
\vfill
\begin{Abstract}
Since magnetic fields in galaxies, galactic clusters and 
even  void regions are observed, theoretical attempts
to explain their origin are strongly motivated.
It is interesting to consider that inflation is responsible for
the origin of the magnetic fields as well as the density perturbation.
However, it is known that inflationary magnetogenesis suffers from
several problems. We explore these problems by using a specific
model, namely the kinetic coupling model, and show how the model is constrained.
Model independent arguments are also discussed.
\end{Abstract}
\vfill
\begin{Presented}
The 10th International Symposium on Cosmology and
Particle Astrophysics (CosPA2013),\\
Honolulu, USA, November 12-15, 2013
\end{Presented}
\vfill
\end{titlepage}
\def\thefootnote{\fnsymbol{footnote}}
\setcounter{footnote}{0}
%

\section{Introduction}

Various magnetic fields are known to exist in the universe.
In the context of cosmology, magnetic fields in galaxies,
galactic clusters and void regions are important.
The magnetic fields in galaxies and clusters are 
$B_{\rm gal} \sim10^{-6}$G~\cite{Wielebinski:2005,Beck:2012}.
The void magnetic field
is reported to be stronger than $B_{\rm void}\gtrsim 10^{-15}$G 
with an uncertainty of a few orders based on blazar observations~\cite{Neronov:1900zz, Tavecchio:2010mk, Essey:2010nd, Finke:2013bua}.
However, the origin of these magnetic fields is a long-term and outstanding
problem in astrophysics and cosmology.

The candidates of the mechanism that generates the magnetic fields
are divided into two main classes~\cite{Giovannini:2003yn,Kandus:2010nw,Durrer:2013pga}. One class includes astrophysical processes
which exploit plasma motions to produce magnetic fields in 
comparatively local regions while it may be difficult for these mechanisms to work in void regions.
The other consists of cosmological processes which generate magnetic fields
spread over the universe in the very early universe.
The magnetic field produced by the latter class of models 
can  directly dilute into the void magnetic field 
and also seed the galactic and cluster magnetic fields 
if the strength is sufficient.
The scenario of the primordial magnetic field  
naturally explains the hierarchy between
$B_{\rm gal}$ and $B_{\rm void}$ because the adiabatic compression
and the dynamo mechanism may amplify it in galaxies and clusters
while the magnetic field is expected to dilute 
due to the cosmic expansion in void regions. 
However, the primordial magnetic field is constrained by CMB observations as 
$B_{p} \lesssim 10^{-9}$G (see, e.g. \cite{Yamazaki:2012pg}, and references therein).
Here, we focus on the magnetic fields with a primordial origin,
especially an inflationary origin.

Inflation is a widely accepted paradigm of 
the very early universe and it can produce cosmological perturbations
from quantum fluctuations. Since the initial density perturbation
which seeds the large scale structures observed in the present universe
originates from inflation, it is an attractive idea that the observed
magnetic fields are also attributed to inflation.
Although many models in which magnetic fields are generated
during inflation are proposed so far~\cite{Turner:1987bw,Ratra:1991bn,Garretson:1992vt,Davis:2000zp,Finelli:2000sh,Ferreira:2013sqa}
, these ``inflationary magnetogenesis" models suffer from several problems.
It is known that the strong couling problem, the back reaction problem 
and the curvature perturbation problem spoil inflationary magnetogenesis models~\cite{Demozzi:2009fu,Fujita:2012rb,Suyama:2012wh,Barnaby:2012xt,Fujita:2013qxa}.
In the following section, we will explain these problems.

Inflationary magnetogenesis targets the magnetic field that is stronger than
the blazar lower bound $B_{\rm void}\gtrsim 10^{-15}$G
because the void magnetic field is not amplified after reheating and reflects
the primordial amplitude.
\footnote{See, however, ref.~\cite{Campanelli:2007tc,Saveliev:2013uva}
in which the inverse cascade is discussed.}
Although $10^{-15}$G is very weak in comparison with, for example, earth's magnetism
($\sim 0.2-0.7$G), an remarkably strong magnetic field 
at the end of inflation is required. It is because the magnetic
field dilutes in proportion to $a^{-2}$ in the expanding universe.
Furthermore, on super horizon scales during inflation, the electric field
is stronger than the magnetic field which has to grow rapidly against
the $a^{-2}$ dilution. As we will see below, this extremely strong
electric field makes magnetogenesis difficult.

\section{Sketch of inflationary magnetogenesis}

The basics of inflationary magnetogenesis can be understood
by reviewing a model. Let us sketch the kinetic coupling model
(or $IFF$ model)~\cite{Ratra:1991bn}
as an example.
The model action is
\begin{equation}
S = \int {\rm d}\eta {\rm d}^3 x \sqrt{-g}
\left[
-\frac{1}{4} I^2(\eta) F_{\mu\nu}F^{\mu\nu}
\right],
\quad
\left(
F_{\mu\nu}\equiv \partial_\mu A_\nu - \partial_\nu A_\mu
\right),
\label{Model Action}
\end{equation}
where $\eta$ is the conformal time, $A_\mu$ is a gauge field and $I(\eta)$ is originally considered
as a function of a scalar field but is treated as a function of time. To solve the EoM of $A_\mu$ analytically,
$I(\eta)$ is usually assumed as
\begin{equation}
I(\phi)=
\left\{
\begin{array}{cc}
 \left(\eta/\eta_\ff \right)^n &  (\eta<\eta_\ff)\\
 1 & (\eta\ge\eta_\ff)
\end{array}\,, \right.
\label{I}
\end{equation}
where ``f'' denotes the end of inflation and $n$ is a constant.
Without a time variation of $I(\eta)$, no fluctuation of $A_\mu$
would not be generated because of the conformal invariance~\cite{Turner:1987bw}.
The EoM of $A_i$ is given by
\begin{equation}
\left[\partial_\eta^2 +k^2-\frac{n(n-1)}{\eta^2}\right](I\mcA_k)=0
\,,
\label{EoM of A}
\end{equation}
where $\mcA_k$ is the mode function of $A_i$ expanded by the polarization
vectors.
If $n<0$, $I(\eta)\ll 1$ during inflation and loop effects due to
the coupling to charged fermions cannot be ignored.
Then a reliable calculation is hardly done.
It is known as the strong coupling problem~\cite{Demozzi:2009fu}.
Thus we choose $n>0$ and obtain the solution on super-horizon scale:
\begin{equation}
| I\mcA_k(\eta) | = 
\frac{\Gamma(n-1/2)}{\sqrt{2\pi k}}\left( \frac{-k\eta}{2}\right)^{1-n}
,
\quad \left(-k\eta \ll 1,\  n>\frac{1}{2} \right).
\label{Sol of A}
\end{equation}
In the expanding universe, the power spectra of electromagnetic fields are given by
\begin{equation}
\mcP_E (\eta,k) \equiv \frac{k^3 |\partial_\eta \mcA_k|^2}{\pi^2 a^4},\qquad
\mcP_B (\eta,k) \equiv \frac{k^5 |\mcA_k|^2}{\pi^2 a^4}.
\label{P of EB}
\end{equation}
It should be noted that the magnetic field is diluted in proportion to $a^{-2}$.
Substituting eq.~\eqref{Sol of A} into eq.~\eqref{P of EB}, one finds
that the resultant magnetic field at present is
\begin{equation}
\mcP^{1/2}_B (\eta_{\rm now},k)
=
\frac{\Gamma(n-\frac{1}{2})}{2^{\frac{3}{2}-n}\pi^{\frac{3}{2}}}
(a_\ff H)^{n-1} k^{3-n}
\sim
10^{23n-80} {\rm G}\times
\left( \frac{\rho_\inf^{1/4}}{10^{16}\GeV} \right)^{n-1}
\left( \frac{k}{1{\rm Mpc}^{-1}} \right)^{3-n},
\label{current B}
\end{equation}
in the case of the instant reheating. Here, $\rho_\inf$ is the inflation energy scale.
Therefore  $n \gtrsim 3$ is necessary to produce the magnetic field
with the sufficient strength, $B(\eta_{\rm now}) > 10^{-15}$G
at $1$ Mpc scale.

\section{Problems}

\subsection{Back reaction problem}

In the previous section, we assume that inflation continues
regardless of the generation of the electromagnetic fields.
However, if the energy density of the electromagnetic fields
overtakes that of inflaton, the inflation dynamics and/or
the generation of electromagnetic fields are altered~\cite{Demozzi:2009fu}.
This is the so-called back reaction problem.

Before evaluating the electromagnetic energy density,
it is important to realize that,
on super-horizon scales, the electric field is much stronger than
the magnetic field in the kinetic coupling model:
\begin{equation}
\left|
\frac{\mcP_E}{\mcP_B}
\right|
=
\left|
\frac{\partial_\eta \mcA_k}{k \mcA_k}
\right|^2
\simeq
\frac{1}{|k\eta|^2}
=e^{2N_k}\gg1,
\qquad ({\rm super\ horizon})
\end{equation}
where $N_k \equiv -\ln|k\eta|$ is the e-folding number of $k$ mode.
Thus we can focus on the electric field.
Its energy density at the end of inflation is given by
\begin{equation}
\rho_\em(\eta_\ff) \simeq \frac{1}{2} \int^{a_\ff H}_{k_\mi}\frac{\dd k}{k}
\mcP_E (\eta_\ff, k)
\simeq
H^4
\left[\frac{e^{2(n-2)N_\tot}-1}{2n-4}\right]
\,,
\end{equation}
where $N_\tot$ is {\it not} the total e-folds of inflation {\it but}
the total e-folds of magnetogenesis (i.e. the time duration where
$I(\eta)\propto \eta^n$) and 
$k_\mi$ is the mode which exits the horizon at the onset of magnetogenesis.
$H$ is the Hubble parameter during inflation.
Note we drop a numerical factor for simplicity.
One can see that for $n>2$, $\rho_\em$ becomes huge due to the IR contribution.

Demozzi et al.~\cite{Demozzi:2009fu} show that by requiring
$\rho_\em < \rho_\inf$, the magnetic field produced in the kinetic coupling
model with the power-law kinetic function, $I(\eta)\propto \eta^n$, cannot exceed $10^{-32}$G
today.
\footnote{They assume $N_\tot=75$ and $H=10^{-6}\Mpl$.}
It is far smaller than the observational lower bound.
Although their result is striking, it does not mean
inflationary magnetogenesis is generally excluded
because their analysis is based on the specific model.
In ref.~\cite{Fujita:2012rb}, nevertheless,
the authors conduct a model independent argument
in which the strong coupling problem and the back reaction problem
are taken into account. They derive an universal upper bound on
the inflation energy scale:
\begin{equation}
\rho^{1/4}_{{\rm inf} }
 <
 6 \times 10^{11} \GeV 
 \left( \frac{B(\eta_{\rm now})}{10^{-15}G} \right)^{-2}.
\label{eq:main result}
\end{equation}
Therefore the back reaction problem implies that 
low energy inflation is favored.
In addition, this constraint can be translated into the bound
on the tensor-to-scalar ratio:
$r<10^{-19}(B/10^{-15}{\rm G})^{-8}$.
Thus if the background gravitational waves are detected in the future,
inflationary magnetogenesis is excluded.

\subsection{Curvature perturbation problem}

The curvature perturbation problem refers that inflationary magnetogenesis
can be constrained due to the curvature perturbation
induced by the generated electromagnetic fields~\cite{Barnaby:2012xt}.
The electromagnetic fields produced during inflation behave
as isocurvature perturbations and source
the adiabatic perturbation~\cite{Suyama:2012wh,Fujita:2013qxa}:
\begin{equation}
\zeta^\em(t,\bm{x}) = -\frac{2H}{\epsilon \rho_{\rm inf}} \int^t_{t_0} dt' \rho_\em (t',\bm{x}),
\label{zeta}
\end{equation}
where $t$ is the cosmic time and $\epsilon$ is the slow-roll parameter.
Regarding the curvature perturbation, 
not only the amplitude of the power spectrum $\mcP_\zeta$ but also 
the non-linarity parameters, $f_{\rm NL}$ and $\tau_{\rm NL}$,
are observationally measured.
Then those parameters induced by the electromagnetic fields
should not exceed the observed values:
\begin{equation}
\mcP^\obs_\zeta > \mcP^\em_\zeta,
\qquad
f_{\rm NL}^\obs > f_{\rm NL}^\em,
\qquad 
\tau_{\rm NL}^\obs > \tau_{\rm NL}^\em.
\quad 
\end{equation}

Considering $\mcP^\obs_\zeta > \mcP^\em_\zeta$ in a model independent way,
the authors in ref.~\cite{Suyama:2012wh} put the lower bound on
the inflation energy scale:
\begin{equation}
\rho_\inf^{1/4} > 3\times 10^{13}\GeV 
 \left( \frac{B(\eta_{\rm now})}{10^{-15}G} \right)^{1/2}.
\end{equation}
Apparently, combined with eq.~\eqref{eq:main result},
this constraint eliminates inflationary magnetogenesis models in general.
Nonetheless it should be noted in ref.~\cite{Suyama:2012wh} the authors assume
that inflation is single slow-roll, the correlation length of 
the void magnetic field is $1$ Mpc at present and the amplitudes of $\mcP_\zeta$ of the minimal scale of inflation is same as that of the CMB scale,
$\mcP_\zeta(k_{\rm CMB})=\mcP_\zeta(k_{\rm max})$. 

On the other hand, without these assumptions, $\mcP^\em_\zeta, f_{\rm NL}^\em$ and $\tau_{\rm NL}^\em$ are explicitly calculated 
and compared with the Planck result~\cite{Ade:2013uln}
 under the framework of the kinetic coupling model in ref.~\cite{Fujita:2013qxa}.
Interestingly, it is found that the constraint from $\tau_{\rm NL}$ is the
most stringent in the single slow-roll inflation case
while the bound from the back reaction problem become the tightest when the single slow-roll assumption is relaxed (see fig.1).
Unfortunately, in both cases, the allowed strength of the magnetic field
is far smaller than the observational lower limit. 

\begin{figure}[htb]
  \hspace{-2mm}
  \includegraphics[width=75mm]{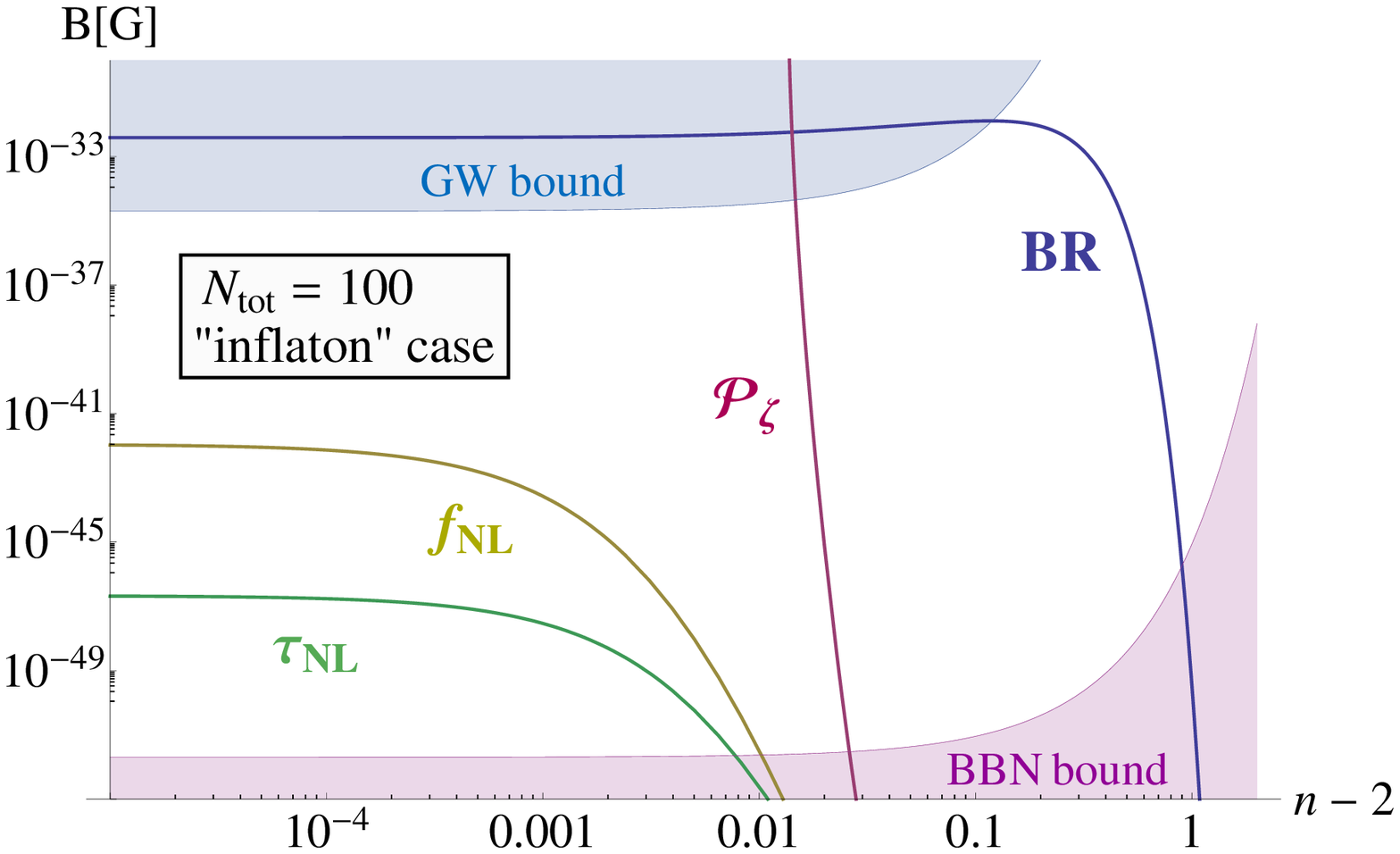}
  \hspace{3mm}
  \includegraphics[width=75mm]{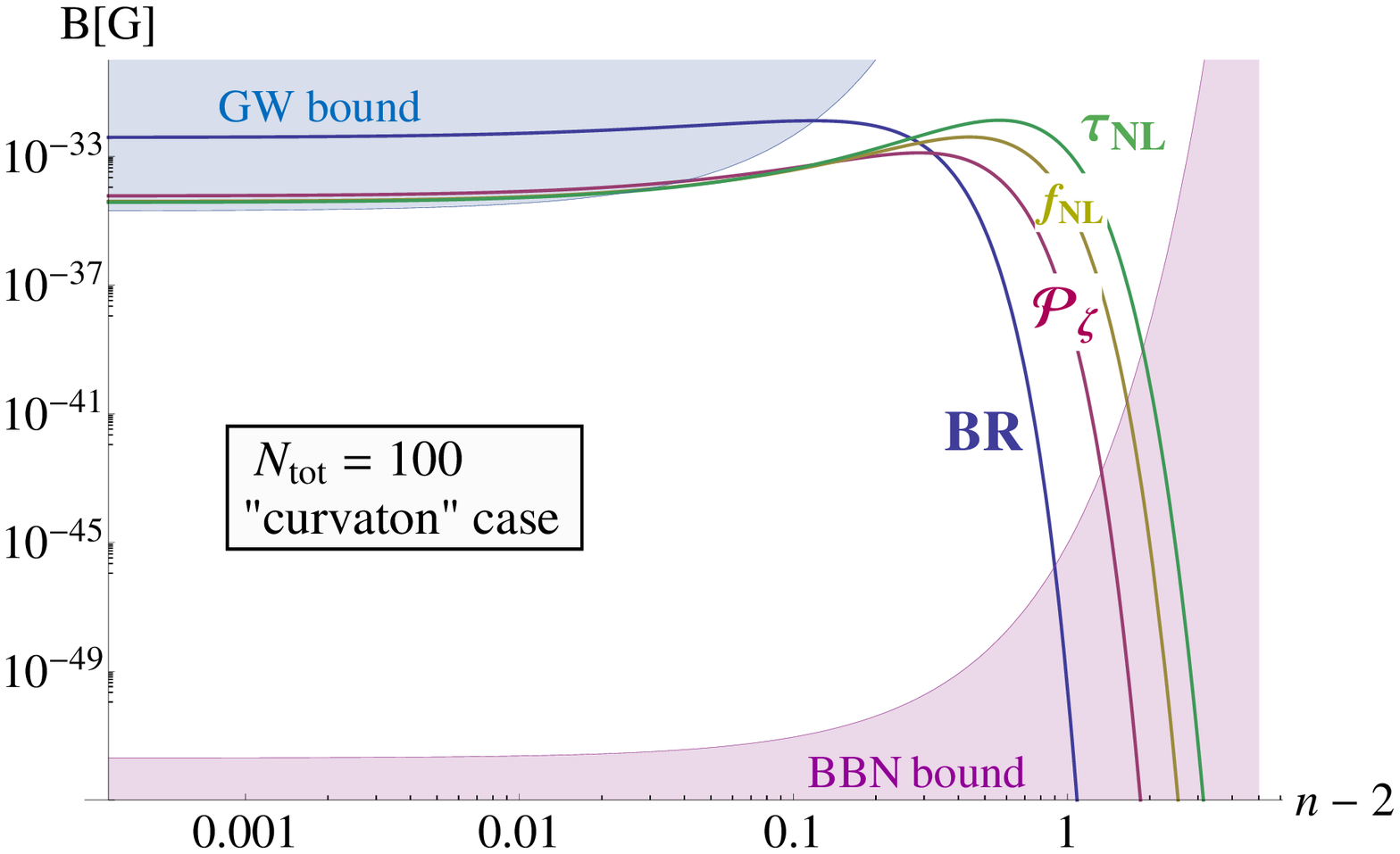}
 \caption
 {The upper limit of the current strength of the magnetic field
  for $n\ge2$.
 in the kinetic coupling model.
 It is assumed that
 inflaton generates all observed curvature perturbation
 in the left panel while that assumption is relaxed and instead 
 $\epsilon = 10^{-2}$ is adopted in
 the right panel. 
 The total duration of the electromagnetic field generation is set as 
 $N_\tot = 100$.
 The shaded regions represent the restriction from
 gravitational wave (blue) and big bang nucleosynthesis (red), respectively.
 }
 \label{fig:B}
\end{figure}

\section{Summary and discussion}

Since the magnetic fields in the universe
are observed and their properties are constrained
($B_{\rm gal} \sim10^{-6} {\rm G}, B_{\rm void}\gtrsim 10^{-15}$G),
theoretical attempts to explain their origin  are strongly motivated.
However, in spite of longstanding efforts and
numerous papers, a successful quantitative model of 
magnetogenesis is not yet established.

In this paper, we explore inflationary magnetogenesis
where the electromagnetic fields are generated during inflation.
The idea that inflation produces the primordial magnetic field
as well as the density perturbation looks natural.
However, as we discussed above, inflationary magnetogenesis
suffers from several problems
and no promising model is known so far.

To determine whether inflationary magnetogenesis is possible
or not, two ways can be considered. One is building a viable model
and explicitly proving its existence. The other is
conducting a model independent argument which generally 
constrains the possibility or gives a guidance for model building.
As the general discussion of the strong coupling and back reaction 
problem~\cite{Fujita:2012rb} implies that
low energy inflation is favored for magntogenesis,
a new general argument will provide a novel insight.
For example, it seems that a model independent argument of the curvature perturbation problem without the assumptions can be made.
\footnote{See ref.~\cite{TFSY}}

Note, we presume that the void magnetic field is generated purely during inflation and no additional amplification occurs.
However, there is a chance that the magnetic field is amplified
during reheating era or its dilution due to the cosmic expansion
is partially compensated by the inverse cascade.
Therefore even if {\it pure} inflationary magnetogenesis is excluded,
the inflationary origin of the cosmic magnetic field
combined with post-inflation dynamics may be viable.

\newpage
\Acknowledgements
This work was supported by the World Premier International
Research Center Initiative
(WPI Initiative), MEXT, Japan. T.F. acknowledges the support by Grant-in-Aid
for JSPS Fellows No.248160.

\end{document}